\documentclass[%
 reprint,
 amsmath,amssymb,
 aps,
]{revtex4-2}

\usepackage{graphicx}
\usepackage{dcolumn}
\usepackage{bm}
\usepackage{hyperref}

\begin{document}

\preprint{APS/123-QED}

\title{Failure of the Crystalline Equivalence Principle for Weak Free Fermions}

\author{Daniel Sheinbaum}
 \affiliation{Division of Applied Physics, CICESE 22860, Ensenada, BC, Mexico.}
 \email{daniels@cicese.mx}
 
\author{Omar Antol\'in Camarena}
 \affiliation{Instituto de Matematicas, National Autonomous University of Mexico, 04510 Mexico City, Mexico.}
\email{omar@matem.unam.mx}

\date{\today}

\begin{abstract}
Interacting crystalline SPT phases were first classified by Thorngren and Else through the crystalline equivalence principle, suggesting that a spatial group symmetry can be treated as if it were an internal symmetry group. Using techniques from topology we elucidate how this principle holds for one of the proposals for the interacting bosonic case and yet fails for weak free fermion phases. Last we show how a variant of the principle does hold for strong crystalline free fermion phases, showing that it is not necessary for the cohomology theory to be of Borel type.
\end{abstract}

\maketitle

\section{Introduction}\label{sec:Introduction}
There are may proposals for the classification of symmetry protected topological (SPT) phases with internal symmetry group $H$ \cite{Wen-SPT}, \cite{Kapustin-SPT}, \cite{Gaiotto-SPT}, \cite{Xiong-SPT}. These proposals have different strengths, assumptions and drawbacks. At the same time, a notion of crystalline topological phase of matter was first discussed for free fermions in \cite{Fu-Crystalline}. Using $\mathit{K}$-theory, \cite{Freed-Moore} classified free fermion weak crystalline phases. There was great interest in combining both of the above notions into what is known as a crystalline SPT phase. In a remarkable proposal, Thorngren and Else \cite{Thorngren-Else} suggested a \textit{crystalline equivalence principle} (CEP), which says you can treat spatial symmetries as if they were internal. This allowed them to employ the group cohomology classification of \cite{Wen-SPT} to construct a classification of crystalline SPT phases. However, to our knowledge, the consistency of this principle when applied to weak free fermions has never been checked! This is relevant because, unlike the interacting case, we have a general consensus on what classifies free fermions \cite{Freed-Moore}. We will give a brief argument explaining why the principle holds for the group cohomology proposal and then show why it fails for the $K$-theory classification of weak free fermion phases.

\section{Equivariant Cohomology vs Equivariant K-theory}
We begin by defining the crystalline symmetry group $G$ as is standard
 \begin{equation}\label{eq:Crystalsymmetry}
     1\rightarrow \mathbb{Z}^d\rightarrow G \rightarrow P\rightarrow 1,
 \end{equation}
 where $\mathbb{Z}^d$ is the subgroup of translations and $P$ is known as the point group, giving rotations and reflections. In the proposal of Thorngren and Else \cite{Thorngren-Else} the CEP implies that what classifies $d$-dimensional $G$ Bosonic SPT phases is
\begin{align}
    H^{d+2}(BG; \mathbb{Z}) = H^{d+2}_{P}(T^{d};\mathbb{Z}),
\end{align}
where $BG$ is known as the classifying space of $G$ \cite{Hatcher-Alg-Top}. The group that appears on the right, $H^{d+2}_{P}(T^{d};\mathbb{Z})$ is what is known as $P$-equivariant cohomology where the group action of $P$ on the real space unit cell torus $T^d$ is determined by eq. (\ref{eq:Crystalsymmetry}).

Now, one can define the internal symmetry group $H$ in a similar fashion:
 \begin{equation}\label{eq:Internalsymmetry}
     1\rightarrow U(1)\rightarrow H \rightarrow Q\rightarrow 1,
 \end{equation}
 where $U(1)$ is charge (particle number) conservation and $Q$ is the parity, time-reversal or charge conjugation symmetry group. We note that for fermions or more specifically superconductors, the $U(1)$ gets reduced to a $(-1)^F$, known as fermion parity. The classification of $H$-SPT phases of Chen et al \cite{Wen-SPT} is given by the group $H^{d+2}(BH; \mathbb{Z})$, thus, when we have both $G$ and $H$ symmetries, the proposal for the general classification for bosons is
 \begin{align}\label{eq:top-crystal-equiv}
    H^{d+2}_{P\times H}(T^d; \mathbb{Z}) = H^{d+2}(B(G\times H);\mathbb{Z}).
\end{align}
We can thus view eq. (\ref{eq:top-crystal-equiv}) as a topological expression for the CEP proposed in \cite{Thorngren-Else}. Note that $G$ and $H$ play the same role in the topological formula.
\section{Free Fermions}\label{sec:free-fermions}
Where as for the interacting case one only has proposals (sometimes called ansatz) for the classification of SPT phases, for free fermions there is more or less a general consensus as to the rigorous classification of crystalline phases \cite{Freed-Moore}, and using the above notation they are classified using various flavors of $\mathit{K}$-theory depending on the symmetry class (represented by a superindex $Q$):
\begin{align}
    \mathbb{K}^{Q,0}_{P}(\mathbb{T}^d) = \mathit{K}^{Q,d}_{P}(T^d),
\end{align}
On the left-hand side, the Brillouin torus $\mathbb{T}^d$ appears, while the right-hand side has the real torus, and the equality comes from crystalline $T$-duality \cite{Thiang-Crystallographic-1},\cite{Thiang-Crystallographic-2} which relates two different kinds of $K$-theory on the two tori. On the left side we have twisted $K$-theory in the sense of \cite{Freed-Moore} (with twists elided from the notation), \cite{FM-Gomi} and on the right we have the ordinary (sometimes twisted) $K$-theory i.e. $K$, $KO$ and $KSP$. We have considered the $T$-dual version in order to clarify the differences between the roles played by $G$ and $H$.  Note that this \emph{is not} $Q$-equivariant $\mathit{K}$-theory on either side. For example in time-reversal symmetry class $AII$ and $P = 1$, on the left we have $\mathbb{K}^{Q,0}_{P} = KR^{-4}$ and on the left $\mathit{K}^{Q,d} = \mathit{KO}^{d-4}$. In general, depending on $Q$ and eq. (\ref{eq:Internalsymmetry}) we have either $\mathit{KO}^{d-n}$ or $\mathit{K}^{d-n}$ for some $n$, depending on the AZ symmetry classes as given in the standard table of topological insulators and superconductors \cite{Kitaev}. First let us emphasize that in the standard free fermion $K$-theoretic classification $H$ does not appear in its entirety, instead one usually employs $H$ (ignoring the $U(1)$ charge conservation symmetry) to obtain a $Q$-action on the space of single particle Hamiltonians i.e. those which anticommute/commute with different antiunitary/unitary representations of $Q$. This will become very relevant later on. Also note that $G$ and $H$ play essentially different roles: the choice of $H$ is equivalent to a choice of $Q$ and a $Q$-action, which selects the flavor of $K$-theory to use (either real or complex) and with which shift (the $n$ above); and $G$ is equivalent to a pair $(T^d, P)$ together with a $P$-action on both $T^d$ and $K^d$ so that it instead specifies the group the theory is equivariant with respect to, its action on $T^d$ is part of the input data to compute $P$-equivariant $K$-theory.

Now, how do we formulate the CEP in topological terms for free fermions? What made the principle true in eq. (\ref{eq:top-crystal-equiv}) was that $P$-equivariance was the same as using the \textit{Borel construction} $T^d\times_{P}EP = BG$ and that $BG$ and $BH$ appear on equal footing on eq. (\ref{eq:top-crystal-equiv}). Indeed, various proposals employ this fact \cite{Freed-Hopkins-Spatial}, \cite{Arun-SPT}, \cite{Xiong-SPT} so that in the same spirit a free fermion CEP in $\mathit{K}$-theory would be a statement of the form ``$\mathit{K}^{Q,d}_{P}(T^d) = \mathit{K}^{d}(T^d\times_{P}EP\times BQ)$''. However, this statement is false on both counts! On the one hand, $P$-equivariant $\mathit{K}$-theory is in many cases \emph{not} ordinary $\mathit{K}$-theory of the Borel construction i.e. \emph{in general}
\begin{equation}\label{eq:failure}
  \mathit{K}^{Q,d}_{P}(T^d) \neq \mathit{K}^{Q,d}(T^d\times_P EP).
\end{equation}
There is a relation between the two groups given by the Atiyah-Segal completion theorem \cite{Atiyah-Completion} but it generally fails to be an equality. On the other hand, the situation is worse for the internal symmetry subgroup $Q$, as not only is $\mathit{K}^{Q,d}(T^d)\neq \mathit{K}^{d}(T^d\times BQ)$ but because of the antiunitarity of some of the symmetries that $Q$ can represent, there is no Atiyah-Segal completion like result to our knowledge. These differences in both cases do not magically cancel out and $H$ and $G$ play a very different role when it comes to $\mathit{K}$-theory.

As suggested by a referee, this argument is clearer in the case of a nonsymmorphic group $G$ with no $G$-fixed points on $\mathbb{R}^d$ (such as the group $\mathsf{pg}$ in $d=2$, containing glide reflections).  In such examples (\ref{eq:failure}) is \emph{false}, that is, an equality to Borel equivariant $K$-theory does hold:
\begin{equation}\label{eq:nonsymmorphic}
  \mathit{K}^{Q,d}_{P}(T^d) = \mathit{K}^{Q,d}(T^d \times_P EP)
\end{equation}
(where there is a twist on the left side due to the projective action of $P$ on Hilbert space, which we omit from the notation \cite{Freed-Moore}, \cite{Gomi-Twists}). Thus, for this class of examples, there is no concern over the two differences magically cancelling, since there is only one difference; our $K$-group is indeed Borel with respect to $P$ yet it is never Borel with respect to any nontrivial choice of $Q$! They cannot be equivalent if one can be Borel and the other never is. To be complete note that to get to (\ref{eq:nonsymmorphic}) we go through a $\mathit{K}$-homology formulation, using the Baum-Connes isomorphism and $G$-equivariant Poincar\'{e} duality \cite{Thiang-Crystallographic-1}, \cite{Thiang-Crystallographic-2}.

\section{Word of Caution to the wise about formulating the equivalence}
Let $GP(d)$ denote the set of gapped Hamiltonians in $d$ dimensions and let $GP(H,G)$ be the set of SPT phases with crystalline group $G$ and internal symmetry $H$. The symmetry groups act on the space of Hamiltonians and systems possesing those symmetries correspond to Hamiltonians which are fixed under both actions. So, a tautological CEP one could formulate is one of the form:
\begin{equation}
   GP(H,G)\equiv  \pi_{0}(GP(d)^{G\times H}),
\end{equation}
which trivially holds by the very definition of $G$ and $H$ being symmetry groups. This statement seems to have $G$ and $H$ playing the same role, since we are looking at Hamiltonians which commute with both types of symmetry and hence it would seem as a sort of CEP. However, the notation hides the main fact that distinguishes between the $G$ and $H$ a priori, which is the different group actions. This formulation of the CEP is clearly tautological, and thus not worth much, in contrast to (\ref{eq:top-crystal-equiv}) which makes calculation feasible. In the case of the current proposals for an interacting classification like group cohomology \cite{Thorngren-Else}, \cite{Freed-Hopkins-Spatial}, \cite{Arun-SPT}, their power and beauty stem from the fact that the different actions disappear and only $B(G\times H)$ remain, and thus one obtains a CEP with physical and mathematical content.

\section{CEP for strong Free fermion phases}
Let us now instead consider the subset of strong free fermion phases \cite{Kitaev}, \cite{Wen-FF}, which in the $T$-dual formulation are those that survive mapping the real torus $T^d$ to a point i.e.
\begin{align}
    \mathit{K}^{Q,d}_P(\cdot) = [\cdot, (K^{d})^{Q\times P}].
\end{align}

Because the space is a point, all that remains of the point group symmetry is what it does to the flavor of $\mathit{K}$-theory, precisely the same role that $Q$ plays! Thus, we have found a CEP for a subset of phases, namely the strong ones. This CEP is similar to the ones found for strong interacting SPT phases in \cite{Freed-Hopkins-Spatial} and \cite{Arun-SPT}.

This example raises a key point: It is \textbf{not} necessary for the cohomology theory that classifies either weak or strong, interacting or non-interacting phases to be of Borel type in order to have a CEP; e.g. $\mathit{K}$-theory is not of Borel type but nevertheless we have a CEP for strong non-interacting phases. Or as a hypothetical example, a CEP for weak free fermion phases could have held if we had the following non-Borelian equivalence
     ``$\mathit{K}_{P}^{Q,d}(T^d) = \mathit{K}^{d}_{Q\times P}(B(U(1)\times \mathbb{Z}^d))$" where we have used that the unit cell torus $T^d$ is precisely $B\mathbb{Z}^d$. Similarly to the case discussed earlier in section \ref{sec:free-fermions}, this is generally false even when fully taking into account the $U(1)$ symmetry in the non-interacting formalism \cite{Wen-FF}.

\section{An ansatz for interacting crystalline phases without a CEP}
At the opposite end of the spectrum, we will now show how one can extend the proposal of Freed and Hopkins \cite{Freed-Hopkins-SPT} to classify interacting crystalline SPT phases, but in a different manner than it was extended in \cite{Freed-Hopkins-Spatial}, so that it does not have a CEP for weak crystalline SPT phases, yet it equals the Freed-Hopkins proposal in the limiting case of no spatial symmetries. Freed and Hopkins propose that interacting SPT phases on a space $Y$ with spatial symmetry group $G$ are given by
\begin{equation}
E_{0,BM}^{hG}(Y) \equiv \pi_0\left((E \wedge Y_+)^{hG}\right).
\end{equation}
Here $E$ is the spectrum they construct in \cite{Freed-Hopkins-Spatial}, which depends on the internal symmetry group $H$ in a Borel fashion, since its definition only involves $H$ through the classifying space $BH$. The notation $(-)^{hG}$ denotes homotopy fixed points, a construction dual to the Borel construction, and given by taking fixed points of the mapping space from $EG$, namely $X^{hG} := \left(Map(EG, X)\right)^G$.

Freed and Hopkins show that their ansatz has a CEP for strong phases when $G$ is the point group (that we called $P$) of the crystalline symmetry group. That is, they show that
\begin{equation}
E_{0,BM}^{hG}(\mathbb{R}^d) = \tilde{E}_{0,BM}(\mathbb{R}^d),
\end{equation}
where $\tilde{E}$ is the spectrum constructed for internal symmetry group $H \times G$.

Imagine that instead of that ansatz one proposed that phases were classified by the actual fixed points, rather than homotopy fixed points, that is by \begin{equation}\label{eqn:non-borel}
    E^G_0(Y) \equiv \pi_0\left((E \wedge Y_+)^G\right).
\end{equation}
This proposal is not Borel with respect to $G$ (though it still is with respect to $H$), since it no longer involves maps from $EG$, and it still agrees with Freed and Hopkins proposal in case $G=1$, i.e., for non-crystalline phases. Comparing with Freed and Hopkins argument for their proposal, shows that this non-Borel variant will not, in general, satisfy a CEP for strong phases. We are not claiming that our proposed formula (\ref{eqn:non-borel}) actually classifies interacting crystalline SPT phases, we merely present it as a potential example of what a classification for interacting crystalline SPT phases that is Borel with respect to internal symmetries but not with respect to spatial ones might look like. We do not know of any a priori reason to discard potential classifications such as this one.

\section{Discussion}
We have seen that there are different possibilities when it comes to having a crystalline equivalence principle (CEP). You could a priori have a CEP for weak phases without the classifying cohomology theory being Borel, or only have one when restricted to strong phases, or not have a CEP at all. If the resulting cohomology theory is of Borel type for both symmetries, then you directly have a CEP.

What we have shown with certainty is that there is no CEP for weak crystalline free fermions though there could have been one in principle and there is a CEP for strong crystalline free fermions, even though $\mathit{K}$-theory is not of Borel type. We have further shown that for interacting systems one can consider different cohomology theories, some that have and some that do not have a CEP and we do not know as of yet a good physical motivation why one should consider one over the other. The group cohomology approach is Borel with respect to both internal and spatial symmetries yielding a CEP, but those which go beyond, using bordism, could perfectly be non-Borel with respect to spatial symmetries. In the original proposal of Thorngren-Else \cite{Thorngren-Else} it is by approximating crystalline SPT phases as topological gauge theories and then gauging the spatial symmetries that the arising classification has a CEP but for more general approaches different possibilities open up.  In a recent example for quantum cellular automata \cite{Else-QCA}, the arising classifying theory seems to be non-Borel with respect to spatial symmetries, yet, as we have shown here, one may or may not have a CEP, so a more careful analysis is required. We believe a broader discussion on the role played by interactions is necessary to zoom in on the right classifying theory in order to figure out what kind of CEP (for weak phases or only for strong phases) is physically realized.

\acknowledgements{We especially thank the referee for improvements on our arguments. We also thank A.~Debray, C.~Krulewski, L.~Stehouwer, N.~Pacheco-Tallaj and D.~Else for useful comments.}


\providecommand{\noopsort}[1]{}\providecommand{\singleletter}[1]{#1}%

\end{document}